\RequirePackage{lineno}

\documentclass[bibtex,twocolumn,prl,showpacs,floatfix,preprintnumbers,amsmath,amssymb,superscriptaddress,letter]{revtex4}

\usepackage{graphicx}% Include figure files
\usepackage{dcolumn}% Align table columns on decimal point
\usepackage{color}
\usepackage{xspace}
\usepackage[tight]{units}
\usepackage{float}
\usepackage{fancyhdr}
\usepackage{tabularx}

\newcommand{\dcp}{\delta_{CP}}
\newcommand{\epsab}{\varepsilon_{\alpha\beta}}
\newcommand{\epset}{\varepsilon_{e\tau}}
\newcommand{\magepset}{|\varepsilon_{e\tau}|}

\newcommand{\numu}{\ensuremath{\nu_{\mu}}\xspace}
\newcommand{\numubar}{\ensuremath{\overline{\nu}_{\mu}}\xspace}

\newcommand{\nue}{\ensuremath{\nu_{e}}}
\newcommand{\nuebar}{\ensuremath{\overline{\nu}_{e}}}

\newcommand{\delcom}{\ensuremath{(\delta_{CP}+\delta_{e\tau})}}

\newcommand{\dmsqtwo}{\Delta m^{2}_{32}}

\newcommand{\dmsqone}{|\Delta m^{2}_{31}|}

\newcommand{\dmsqonenb}{\Delta m^{2}_{31}}
\newcommand{\dmsqonetwonb}{\Delta m^{2}_{21}}

\makeatletter

\newcommand{\Rmnum}[1]{\expandafter\@slowromancap\romannumeral #1@}
\makeatother

\begin{document}

\vspace*{.6cm} 
\title{A search for flavor-changing non-standard neutrino interactions using $\nue$ appearance in MINOS}
\newcommand{\Berkeley}{Lawrence Berkeley National Laboratory, Berkeley, California, 94720 USA}
\newcommand{\Cambridge}{Cavendish Laboratory, University of Cambridge, Madingley Road, Cambridge CB3 0HE, United Kingdom}
\newcommand{\Cincinnati}{Department of Physics, University of Cincinnati, Cincinnati, Ohio 45221, USA}
\newcommand{\FNAL}{Fermi National Accelerator Laboratory, Batavia, Illinois 60510, USA}
\newcommand{\RAL}{Rutherford Appleton Laboratory, Science and Technology
 Facilities Council, Didcot, OX11 0QX, United Kingdom}
\newcommand{\UCL}{Department of Physics and Astronomy, University College London, Gower Street, London WC1E 6BT, United Kingdom}
\newcommand{\Caltech}{Lauritsen Laboratory, California Institute of Technology, Pasadena, California 91125, USA}
\newcommand{\Alabama}{Department of Physics and Astronomy, University of Alabama, Tuscaloosa, Alabama 35487, USA}
\newcommand{\ANL}{Argonne National Laboratory, Argonne, Illinois 60439, USA}
\newcommand{\Athens}{Department of Physics, University of Athens, GR-15771 Athens, Greece}
\newcommand{\NTUAthens}{Department of Physics, National Tech. University of Athens, GR-15780 Athens, Greece}
\newcommand{\Benedictine}{Physics Department, Benedictine University, Lisle, Illinois 60532, USA}
\newcommand{\BNL}{Brookhaven National Laboratory, Upton, New York 11973, USA}
\newcommand{\CdF}{APC -- Universit\'{e} Paris 7 Denis Diderot, 10, rue Alice Domon et L\'{e}onie Duquet, F-75205 Paris Cedex 13, France}
\newcommand{\Cleveland}{Cleveland Clinic, Cleveland, Ohio 44195, USA}
\newcommand{\Delhi}{Department of Physics \& Astrophysics, University of Delhi, Delhi 110007, India}
\newcommand{\GEHealth}{GE Healthcare, Florence South Carolina 29501, USA}
\newcommand{\Harvard}{Department of Physics, Harvard University, Cambridge, Massachusetts 02138, USA}
\newcommand{\HolyCross}{Holy Cross College, Notre Dame, Indiana 46556, USA}
\newcommand{\Houston}{Department of Physics, University of Houston, Houston, Texas 77204, USA}
\newcommand{\IIT}{Department of Physics, Illinois Institute of Technology, Chicago, Illinois 60616, USA}
\newcommand{\Iowa}{Department of Physics and Astronomy, Iowa State University, Ames, Iowa 50011 USA}
\newcommand{\Indiana}{Indiana University, Bloomington, Indiana 47405, USA}
\newcommand{\ITEP}{High Energy Experimental Physics Department, ITEP, B. Cheremushkinskaya, 25, 117218 Moscow, Russia}
\newcommand{\JMU}{Physics Department, James Madison University, Harrisonburg, Virginia 22807, USA}
\newcommand{\LASL}{Nuclear Nonproliferation Division, Threat Reduction Directorate, Los Alamos National Laboratory, Los Alamos, New Mexico 87545, USA}
\newcommand{\Lebedev}{Nuclear Physics Department, Lebedev Physical Institute, Leninsky Prospect 53, 119991 Moscow, Russia}
\newcommand{\LLL}{Lawrence Livermore National Laboratory, Livermore, California 94550, USA}
\newcommand{\LosAlamos}{Los Alamos National Laboratory, Los Alamos, New Mexico 87545, USA}
\newcommand{\Manchester}{School of Physics and Astronomy, University of Manchester, Oxford Road, Manchester M13 9PL, United Kingdom}
\newcommand{\MIT}{Lincoln Laboratory, Massachusetts Institute of Technology, Lexington, Massachusetts 02420, USA}
\newcommand{\Minnesota}{University of Minnesota, Minneapolis, Minnesota 55455, USA}
\newcommand{\Crookston}{Math, Science and Technology Department, University of Minnesota -- Crookston, Crookston, Minnesota 56716, USA}
\newcommand{\Duluth}{Department of Physics, University of Minnesota Duluth, Duluth, Minnesota 55812, USA}
\newcommand{\Ohio}{Center for Cosmology and Astro Particle Physics, Ohio State University, Columbus, Ohio 43210 USA}
\newcommand{\Otterbein}{Otterbein College, Westerville, Ohio 43081, USA}
\newcommand{\Oxford}{Subdepartment of Particle Physics, University of Oxford, Oxford OX1 3RH, United Kingdom}
\newcommand{\PennState}{Department of Physics, Pennsylvania State University, State College, Pennsylvania 16802, USA}
\newcommand{\PennU}{Department of Physics and Astronomy, University of Pennsylvania, Philadelphia, Pennsylvania 19104, USA}
\newcommand{\Pittsburgh}{Department of Physics and Astronomy, University of Pittsburgh, Pittsburgh, Pennsylvania 15260, USA}
\newcommand{\IHEP}{Institute for High Energy Physics, Protvino, Moscow Region RU-140284, Russia}
\newcommand{\Rochester}{Department of Physics and Astronomy, University of Rochester, New York 14627 USA}
\newcommand{\RoyalH}{Physics Department, Royal Holloway, University of London, Egham, Surrey, TW20 0EX, United Kingdom}
\newcommand{\Carolina}{Department of Physics and Astronomy, University of South Carolina, Columbia, South Carolina 29208, USA}
\newcommand{\SDakota}{South Dakota School of Mines and Technology, Rapid City, South Dakota 57701, USA}
\newcommand{\SLAC}{Stanford Linear Accelerator Center, Stanford, California 94309, USA}
\newcommand{\Stanford}{Department of Physics, Stanford University, Stanford, California 94305, USA}
\newcommand{\StJohnFisher}{Physics Department, St. John Fisher College, Rochester, New York 14618 USA}
\newcommand{\Sussex}{Department of Physics and Astronomy, University of Sussex, Falmer, Brighton BN1 9QH, United Kingdom}
\newcommand{\TexasAM}{Physics Department, Texas A\&M University, College Station, Texas 77843, USA}
\newcommand{\Texas}{Department of Physics, University of Texas at Austin, 1 University Station C1600, Austin, Texas 78712, USA}
\newcommand{\TechX}{Tech-X Corporation, Boulder, Colorado 80303, USA}
\newcommand{\Tufts}{Physics Department, Tufts University, Medford, Massachusetts 02155, USA}
\newcommand{\UNICAMP}{Universidade Estadual de Campinas, IFGW-UNICAMP, CP 6165, 13083-970, Campinas, SP, Brazil}
\newcommand{\UFG}{Instituto de F\'{i}sica, 
Universidade Federal de Goi\'{a}s, 74690-900, Goi\^{a}nia, GO, Brazil}
\newcommand{\USP}{Instituto de F\'{i}sica, Universidade de S\~{a}o Paulo,  CP 66318, 05315-970, S\~{a}o Paulo, SP, Brazil}
\newcommand{\Warsaw}{Department of Physics, University of Warsaw, Pasteura 5, PL-02-093 Warsaw, Poland}
\newcommand{\Washington}{Physics Department, Western Washington University, Bellingham, Washington 98225, USA}
\newcommand{\WandM}{Department of Physics, College of William \& Mary, Williamsburg, Virginia 23187, USA}
\newcommand{\Wisconsin}{Physics Department, University of Wisconsin, Madison, Wisconsin 53706, USA}
\newcommand{\deceased}{Deceased.}

\affiliation{\ANL}
%\affiliation{\Athens}
%\affiliation{\Benedictine}
\affiliation{\BNL}
\affiliation{\Caltech}
\affiliation{\Cambridge}
\affiliation{\UNICAMP}
%\affiliation{\CdF}
\affiliation{\Cincinnati}
\affiliation{\FNAL}
\affiliation{\UFG}
\affiliation{\Harvard}
\affiliation{\HolyCross}
\affiliation{\Houston}
\affiliation{\IIT}
\affiliation{\Indiana}
\affiliation{\Iowa}
%\affiliation{\IHEP}
%\affiliation{\ITEP}
%\affiliation{\JMU}
%\affiliation{\Lebedev}
%\affiliation{\LLL}
\affiliation{\UCL}
\affiliation{\Manchester}
\affiliation{\Minnesota}
\affiliation{\Duluth}
\affiliation{\Otterbein}
\affiliation{\Oxford}
\affiliation{\Pittsburgh}
\affiliation{\RAL}
\affiliation{\USP}
\affiliation{\Carolina}
\affiliation{\Stanford}
\affiliation{\Sussex}
\affiliation{\TexasAM}
\affiliation{\Texas}
\affiliation{\Tufts}
\affiliation{\Warsaw}
%\affiliation{\Washington}
\affiliation{\WandM}
%\affiliation{\Wisconsin}

\author{P.~Adamson}
\affiliation{\FNAL}
%\affiliation{\UCL}
%\affiliation{\Sussex}

%\author{C.~Andreopoulos}
%\affiliation{\RAL}
%\affiliation{\Athens}

\author{I.~Anghel}
\affiliation{\Iowa}
\affiliation{\ANL}

%\author{K.~E.~Arms}
%\affiliation{\Minnesota}

%\author{R.~Armstrong}
%\affiliation{\Indiana}

\author{A.~Aurisano}
\affiliation{\Cincinnati}

%\author{T.~H.~Fields}
%\affiliation{\ANL}

%\author{D.~J.~Auty}
%\affiliation{\Sussex}

%\author{S.~Avvakumov}
%\affiliation{\Stanford}

%\author{D.~S.~Ayres}
%\affiliation{\ANL}

%\author{C.~Backhouse}
%\affiliation{\Oxford}

%\author{B.~Baller}
%\affiliation{\FNAL}

%\author{B.~Barish}
%\affiliation{\Caltech}

%\author{P.~D.~Barnes~Jr.}
%\affiliation{\LLL}

\author{G.~Barr}
\affiliation{\Oxford}

%\author{W.~L.~Barrett}
%\affiliation{\Washington}

%\author{E.~Beall}
%\altaffiliation[Now at\ ]{\Cleveland .}
%\affiliation{\ANL}
%\affiliation{\Minnesota}

%\author{B.~R.~Becker}
%\affiliation{\Minnesota}

%\author{A.~Belias}
%\affiliation{\RAL}

%\author{R.~H.~Bernstein}
%\affiliation{\FNAL}

%\author{M.~Betancourt}
%\affiliation{\Minnesota}

%\author{D.~Bhattacharya}
%\affiliation{\Pittsburgh}

%\author{M.~Bhattarai}
%\affiliation{\Texas}
%\affiliation{\Duluth}

\author{M.~Bishai}
\affiliation{\BNL}

\author{A.~Blake}
\altaffiliation[Now at\ ]{Lancaster University, Lancaster, LA1 4YB, UK.}
\affiliation{\Cambridge}
%\author{B.~Bock}
%\affiliation{\Duluth}

\author{G.~J.~Bock}
\affiliation{\FNAL}

%\author{D.~J.~Boehnlein}
%\affiliation{\FNAL}

\author{D.~Bogert}
\affiliation{\FNAL}

%\author{P.~M.~Border}
%\affiliation{\Minnesota}

%\author{C.~Bower}
%\affiliation{\Indiana}

%\author{E.~Buckley-Geer}
%\affiliation{\FNAL}

\author{S.~V.~Cao}
\affiliation{\Texas}

\author{T.~J.~Carroll}
\affiliation{\Texas}

% ...

\collaboration{The MINOS Collaboration}
\noaffiliation

\begin{abstract}
We report new constraints on flavor-changing non-standard neutrino interactions from the MINOS long-baseline experiment using $\nue$ and $\nuebar$ appearance candidate events from predominantly $\numu$ and $\bar{\nu}_\mu$ beams. We used a statistical selection algorithm to separate $\nue$ candidates from background events, enabling an analysis of the combined MINOS neutrino and antineutrino data. We observe no deviations from standard neutrino mixing, and thus place constraints on the non-standard interaction matter effect, $\magepset$, and phase, $\delcom$, using a thirty-bin likelihood fit.
\end{abstract}

\maketitle
\thispagestyle{fancy}

Results from solar, atmospheric, reactor and accelerator experiments~\cite{ref:sk,ref:sno,ref:minos2006,ref:kamland,ref:borexino,ref:k2k,ref:db_reno} demonstrate that neutrinos undergo flavor change as they propagate. This phenomenon is well described by a quantum mechanical mixing of the neutrino flavors. In the standard three-flavor oscillation model, this process can be parameterized by three angles, $\theta_{12},\,\theta_{13},\,\theta_{23}$, and a CP-violating phase, $\dcp$~\cite{ref:pmns}. Electron neutrinos and antineutrinos propagating through matter are subjected to the Mikheyev-Smirnov-Wolfenstein~(MSW) effect~\cite{ref:msw_original}, which arises from forward coherent scattering with electrons in media. While this process itself does not change lepton flavor, the scattering modifies the probabilities for neutrinos to oscillate between flavor states.  

Non-standard interactions (NSI)~\cite{ref:nsi, ref:fried, ref:ohl} that enter the oscillation model permit additional interactions between matter and all neutrino flavors. Analogous to the MSW Hamiltonian, NSI effects can be added as a perturbation to the Hamiltonian associated with three-flavor vacuum neutrino oscillation,
\begin{linenomath}\begin{equation}
H_{mat} = \sqrt{2}G_{F}N_{e}\begin{bmatrix} 1+\varepsilon_{ee} & \varepsilon^{*}_{e\mu} & \varepsilon^{*}_{e\tau} \\ \varepsilon_{e\mu} & \varepsilon_{\mu\mu} & \varepsilon^{*}_{\mu\tau} \\ \varepsilon_{e\tau} & \varepsilon_{\mu\tau} & \varepsilon_{\tau\tau} \end{bmatrix}
\label{eq:nsipt}.
\end{equation}\end{linenomath}
This addition depends upon the matter potential $V = \sqrt{2}G_{F}N_{e}$, where $G_{F}$ is the Fermi coupling constant and $N_{e}$ is the electron density of the traversed material; as well as the $\varepsilon_{\alpha\beta}$ complex coefficients, which indicate the strength of the NSI couplings.

The net impact of adding NSI to our oscillation model yields nine additional free parameters: six $|\epsab|$ terms and three additional CP-violating phases, $\delta_{e\tau}$, $\delta_{e\mu}$, and $\delta_{\mu\tau}$, which couple to the off-diagonal elements in Eq.~(\ref{eq:nsipt}).

Searches for NSI have already been performed on an atmospheric neutrino sample at Super-Kamiokande~\cite{ref:superK}, the $\numu$ and $\numubar$ survival channels at MINOS~\cite{ref:zey,ref:Mann_PRD,Kopp:2010qt,ref:ZeynepThesis}, and $\nue$ and $\nuebar$ appearance channels in MINOS and T2K~\cite{ref:Mann}. Long-baseline accelerator-based experiments offer well-defined propagation lengths and tunable energy spectra that benefit oscillation searches, including investigations of potential NSI phenomena. For this paper, we only consider propagation-induced effects. Both detection and production sources of NSI, as discussed in~\cite{ref:ohl,ref:Kopp}, are being ignored.

MINOS~\cite{ref:minosnim} uses two magnetized steel-scintillator tracking calorimeters placed along the NuMI beamline~\cite{ref:NuMI} to study neutrino oscillation. In the NuMI beam, pions and kaons are produced from interactions between incident 120 GeV protons and a graphite target. Charged particles are subsequently focused by a pair of current-pulsed aluminum horns, and the decays of these particles in a \unit[675]{m} decay pipe yield neutrinos. Remaining hadrons are captured by an absorber at the end of the decay pipe, and muons are removed from the beam by 240 meters of rock that separate the absorber from the detector hall. The neutrino energy spectrum depends upon the configuration of the focusing horns. For this analysis, the reconstructed energy spectrum peaked at around \unit[3]{GeV}. 

The 0.98 kiloton Near Detector (ND) is located \unit[1.04]{km} downstream of the NuMI target at the Fermi National Accelerator Laboratory. It provides observations of the initial composition of the neutrino beam. From Monte Carlo simulations at the ND, we determined that the beam consists of 91.7$\%$ $\numu$, 7.0$\%$ $\numubar$, and 1.3$\%$ ($\nue + \nuebar$) when operated in neutrino mode, and 58.1$\%$ $\numu$, 39.9$\%$ $\numubar$, and 2.0$\%$ ($\nue + \nuebar$) when operated in antineutrino mode~\cite{ref:4th}. The 5.4 kiloton Far Detector (FD), situated in the Soudan Mine in northeastern Minnesota \unit[735]{km} downstream of the NuMI target, registers neutrino interactions and permits searches for oscillation phenomena. 

MINOS has investigated the $\nue$ appearance channel, yielding constraints on both the $\theta_{13}$ and $\delta_{CP}$ mixing parameters~\cite{ref:4th,ref:AdamThesis}. The tightest constraints on $\theta_{13}$ come from reactor neutrino experiments, which have shown the parameter to be non-zero~\cite{ref:db_reno}. By using the MINOS $\nue$ and $\nuebar$ appearance data in conjunction with these reactor results, limits can be placed upon the $\magepset$ parameter. This technique is demonstrated in~\cite{ref:Mann}, which obtained confidence limits on $\magepset$ based upon the MINOS and T2K $\nue$ appearance event rates. In this paper, we improve upon the techniques of~\cite{ref:Mann} by utilizing the observed energy spectra of $\nue$ and $\nuebar$ appearance candidates and a proper treatment of the systematic uncertainties. We also introduce the uncertainties of the other NSI $\epsab$ parameters to the fitting framework. We adopt the same selection criteria and MINOS exposure as~\cite{ref:4th}.
 
A fast matrix multiplication method was introduced to the $\nue$ analysis software to calculate the exact oscillation probabilities, as opposed to using channel-by-channel approximations. In this paper, we show results in terms of $\magepset$ and an effective phase, ($\dcp+\delta_{e\tau})$. The use of this phase was motivated by the MINOS+T2K combined analysis~\cite{ref:Mann}, where it was demonstrated that the $\nue$ appearance probability can be represented in terms of ($\dcp+\delta_{e\tau})$ in the limit of $\dmsqonetwonb/\dmsqonenb\rightarrow0$. Here, we maintain the use of this phase to simplify the presentation of the results even though we have implemented an exact probability calculation that treats all of the oscillation parameters (including NSI) independently. 

 The results presented here are based upon exposures of 10.6~$\times$~10$^{20}$ protons-on-target (POT) in neutrino mode and 3.3~$\times$~10$^{20}$ POT in antineutrino mode. These data were previously analyzed in the first joint $\nue+\nuebar$ appearance search~\cite{ref:4th}.$\newline$
 \indent Events in the $\nue$ and $\nuebar$ charged current (CC) samples were identified using a statistical selection algorithm. Unlike the MINOS $\numu$ and $\numubar$ CC analyses in which events are easily distinguished and charge-sign selected based upon the presence of muon tracks and their curvatures, $\nue$ analyses must address the similar topologies of the $\nue$($\nuebar$) CC signal and neutral current (NC) background. A Library Event Matching (LEM) technique, adopted in previous searches~\cite{ref:4th,ref:TonerThesis,ref:AdamThesis,ref:3rd}, was used to compare the energy depositions of input candidates to libraries of 20 million $\nue$ or $\nuebar$ CC interactions, with 30 million NC interactions in both libraries. Information gathered through this matching process was passed to an artificial neural network that returned a single-valued discriminant, $0.0 < \alpha_{LEM} < 1.0$, for each event. 
 
\begin{figure}[]
    \centering
    \includegraphics[trim = 0mm 0mm 0mm 0mm, clip, width=\columnwidth]{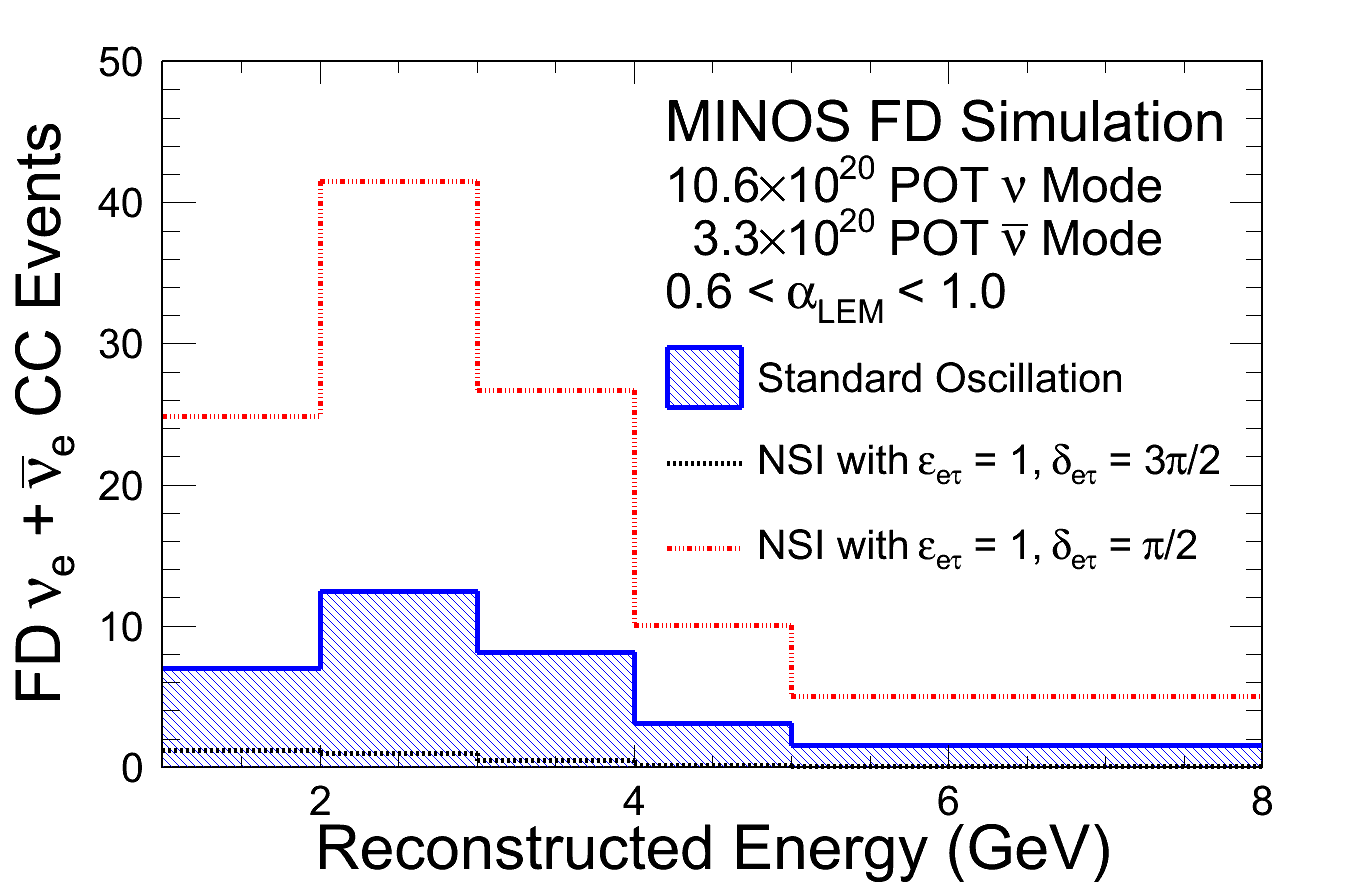}\\
    \caption{Far Detector Monte Carlo $\nue+\nuebar$ CC reconstructed energy distributions shown for standard oscillation and for two illustrative non-standard oscillation cases. Normal mass hierarchy, $\dmsqone = 2.41\times10^{-3} eV^{2}$, $\dcp = 0$, $\theta_{23} = \pi/4$ and $\sin^{2}(2\theta_{13}) = 0.084$ are assumed. Unspecified NSI parameters are set to zero.\label{fig:spec}}
    \end{figure}

  Events with $\alpha_{LEM} > 0.6$, considered sensitive to $\nue$ and $\nuebar$ appearance, constitute our analysis samples. The cut on $\alpha_{LEM}$ was established in~\cite{ref:4th} through the assessment of different binning schemes and the resulting sensitivities to $\theta_{13}$. Assuming a three-flavor neutrino oscillation model that includes the Hamiltonian perturbation from Eq.~(\ref{eq:nsipt}), we probe the $\magepset,\delcom$ parameter space by comparing predicted FD event counts with those observed. A simultaneous fit of the neutrino and antineutrino data is performed using a 30 bin scheme. Each configuration is represented by 15 bins with three divisions of $\alpha_{LEM} > 0.6$ and five divisions that span a reconstructed neutrino energy range of \unit[1-8]{GeV}. Reconstructed energy distributions, shown in Fig.~\ref{fig:spec}, compare three FD Monte Carlo predictions that illustrate standard and non-standard oscillations.
  
\begin{figure}[]
    \centering
    \includegraphics[trim = 0mm 0mm 0mm 0mm, clip, width=\columnwidth]{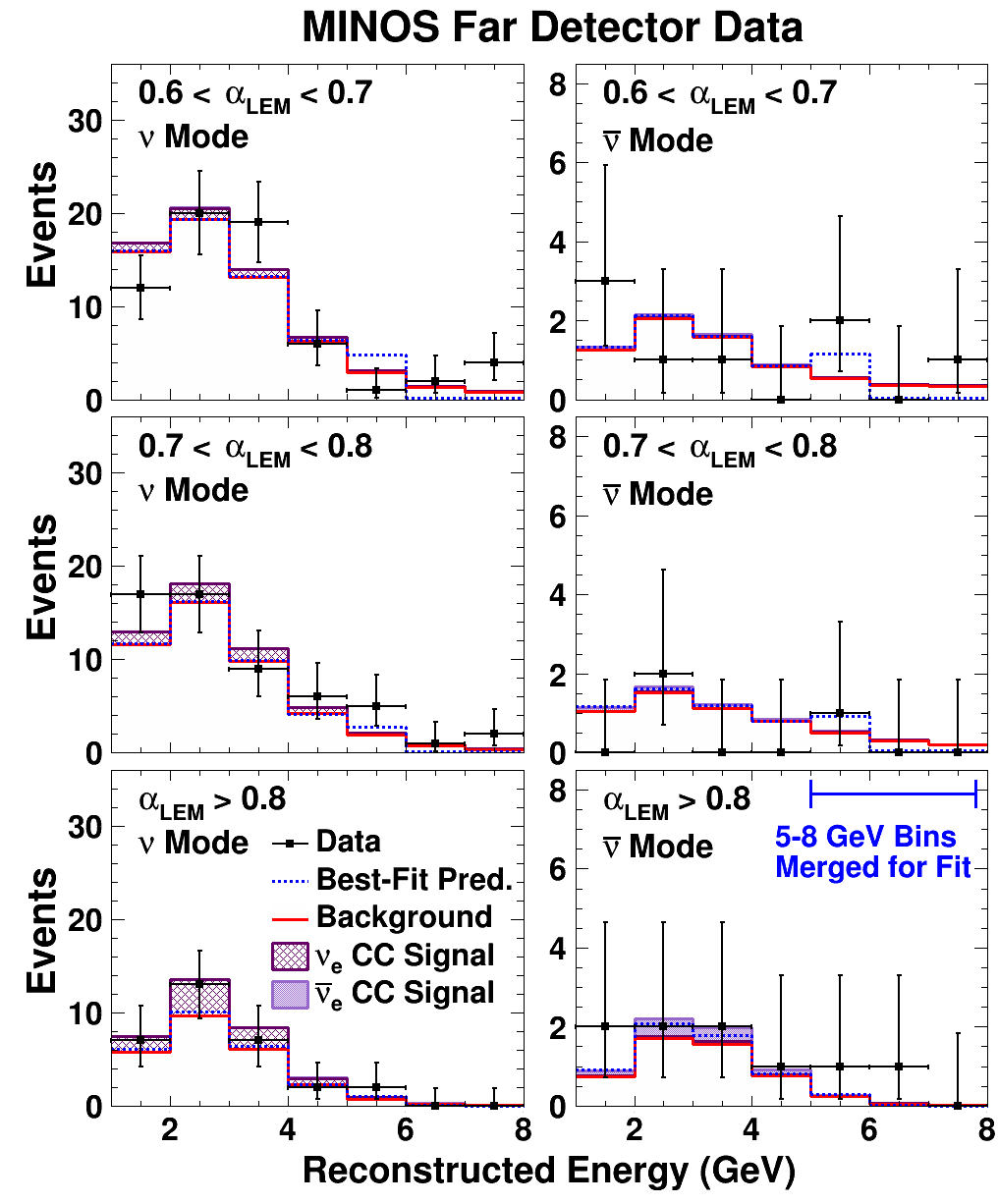}\\
    \caption{The reconstructed energy distributions for three ranges of $\alpha_{LEM}.$ The \unit[5-8]{GeV} region is combined into a single bin for the fit. The best fit that includes systematics is shown assuming normal mass hierarchy, with $\magepset = 0.74$ and $(\dcp+\delta_{e\tau}) = 1.35\pi$. It is overlaid atop the standard oscillation predictions from~\cite{ref:4th}.\label{fig:busy}}
\end{figure} 

Figure~\ref{fig:busy} shows the reconstructed energy distributions for data, the standard oscillation prediction, and the NSI-allowed best-fit prediction. The divisions of $\alpha_{LEM}$, the splitting of neutrino and antineutrino modes, and the reconstructed energy range of the shown histograms are representative of the binning scheme used in the fit.
      
   A two-dimensional fitting framework assessed likelihoods at values of both $\magepset$ and $\delta_{e\tau}$ given a selected neutrino mass hierarchy and $\dcp$. All of the produced likelihood surfaces were represented in terms of ($\dcp+\delta_{e\tau}$), and the contours for each hierarchy were generated by sampling the surfaces for different values of $\dcp$ to determine the most conservative result. Statistical and systematic uncertainties on the FD prediction were taken into account when assessing the contours. In this analysis, the handling of nuisance parameters corresponding to systematic errors and oscillation parameters is identical to the prescription in~\cite{ref:4th} with one notable exception.
 
 To account for the uncertainties on the oscillation parameters, templates were introduced to the fitting framework that treat $\theta_{23}$, $\dmsqtwo$, and five of the NSI $\varepsilon_{\alpha\beta}$ terms as nuisance parameters. The mean values and uncertainties, taken from external data, of the nuisance parameters used in the fit are included in Table~\ref{table:unc}. The templates were made by storing the changes in the FD prediction induced by $\pm1\sigma$ shifts on the selected oscillation parameters in each bin of reconstructed energy and $\alpha_{LEM}$. These templates cover the range of allowed spectra, including those produced through the interplay of the NSI parameters. The stored predictions improved the efficiency of the fit by shortening the amount of time needed to evaluate the likelihood at each point in the parameter space. Central values and limits were taken from \cite{ref:daya} for $\theta_{13}$; from \cite{ref:forero} for the other standard oscillation parameters; from \cite{ref:superK}, for $\varepsilon_{\mu\tau}$; and from  \cite{ref:ohl} for the other $\varepsilon_{\alpha\beta}$ coefficients. 
 
 \begin{table}[]
  \begin{tabularx}{\columnwidth}{l r c c} %{ l l  p{3.75cm} }
        \hline
        \hline
        Parameter & Mean & Uncertainty & Reference\\ \hline
         $\varepsilon_{ee}$   & 0.00 & ($-$4.20,$+$4.20) & \cite{ref:ohl}  \\ 
         $\varepsilon_{e\mu}$ & 0.00 & ($-$0.33,$+$0.33) & \cite{ref:ohl} \\
         $\varepsilon_{\mu\mu}$ & 0.00 & ($-$0.07,$+$0.07) & \cite{ref:ohl} \\
         $\varepsilon_{\mu\tau}$ & 0.00 & ($-$0.01,$+$0.01) & \cite{ref:superK} \\
         $\varepsilon_{\tau\tau}$ & 0.00 & ($-$21.0,$+$21.0) & \cite{ref:ohl}\\ 
         $\dmsqtwo / 10^{-3}$ eV$^{2}$ (NH) & $+$2.404 & ($-$0.068,$+$0.048) & \cite{ref:forero} \\
         $\dmsqtwo / 10^{-3}$ eV$^{2}$ (IH) & $-$2.304 & ($-$0.058,$+$0.048) & \cite{ref:forero} \\
         $\theta_{23}$ (NH) & 0.853 & ($-$0.128,$+$0.033) & \cite{ref:forero} \\ 
         $\theta_{23}$ (IH) & 0.859 & ($-$0.043,$+$0.025) & \cite{ref:forero}\\
         \hline\hline
 \end{tabularx}
 \caption{The mean values and uncertainties placed upon the nuisance parameters in the fit, with respective references shown. Oscillation parameters with mass hierarchy dependent selections are designated with (NH) or (IH) for normal or inverted hierarchy, respectively. We conservatively treat the 90$\%$ C.L. of the NSI parameters as the uncertainties in our fit. Both $\delta_{e\mu}$ and $\delta_{\mu\tau}$ are explicitly set to zero. The inclusion of negative values of $\epsab$ is equivalent to shifting the corresponding phases by $\pi$ radians. We note that through this method the fit spans the full range of how these parameters impact the $\nue$ and  $\nuebar$ appearance rates.} \label{table:unc}
 \end{table}
 
  \begin{table}[]
  \begin{tabularx}{\columnwidth}{l c c} %{ l l  p{3.75cm} }
        \hline\hline
        Parameter & Best Fit (NH) & Best Fit (IH) \\ \hline
         $\varepsilon_{ee}$   & -0.26$\sigma$ & -0.21$\sigma$  \\ 
         $\varepsilon_{e\mu}$ & 0.54$\sigma$ & 0.48$\sigma$  \\
         $\varepsilon_{\mu\mu}$ & 0.00$\sigma$ & 0.57$\sigma$  \\
         $\varepsilon_{\mu\tau}$ & 0.01$\sigma$ & 0.01$\sigma$  \\
         $\varepsilon_{\tau\tau}$ & 0.00$\sigma$ & 0.24$\sigma$ \\ 
         $\dmsqtwo / 10^{-3}$ eV$^{2}$ & 0.00$\sigma$ & 0.00$\sigma$ \\
         $\theta_{23}$ & -0.07$\sigma$ & -0.08$\sigma$ \\ \hline
         $\magepset$ & 0.74 & 0.58 \\
         $\dcp+\delta_{e\tau}$ & 1.35$\pi$ & 1.65$\pi$ \\
         \hline\hline
         
 \end{tabularx}
 \caption{The standard deviations from the central values of the penalty terms as determined by the fit along with the best-fit values of  $|\epset|$ and $\dcp+\delta_{e\tau}$. Normal and inverted hierarchy are designated by (NH) and (IH) respectively.} \label{table:con}
 \end{table}
 
  \begin{figure}[]
   \centering
   \includegraphics[trim = 0mm 20.4mm 0mm 12mm, clip, width=\columnwidth]{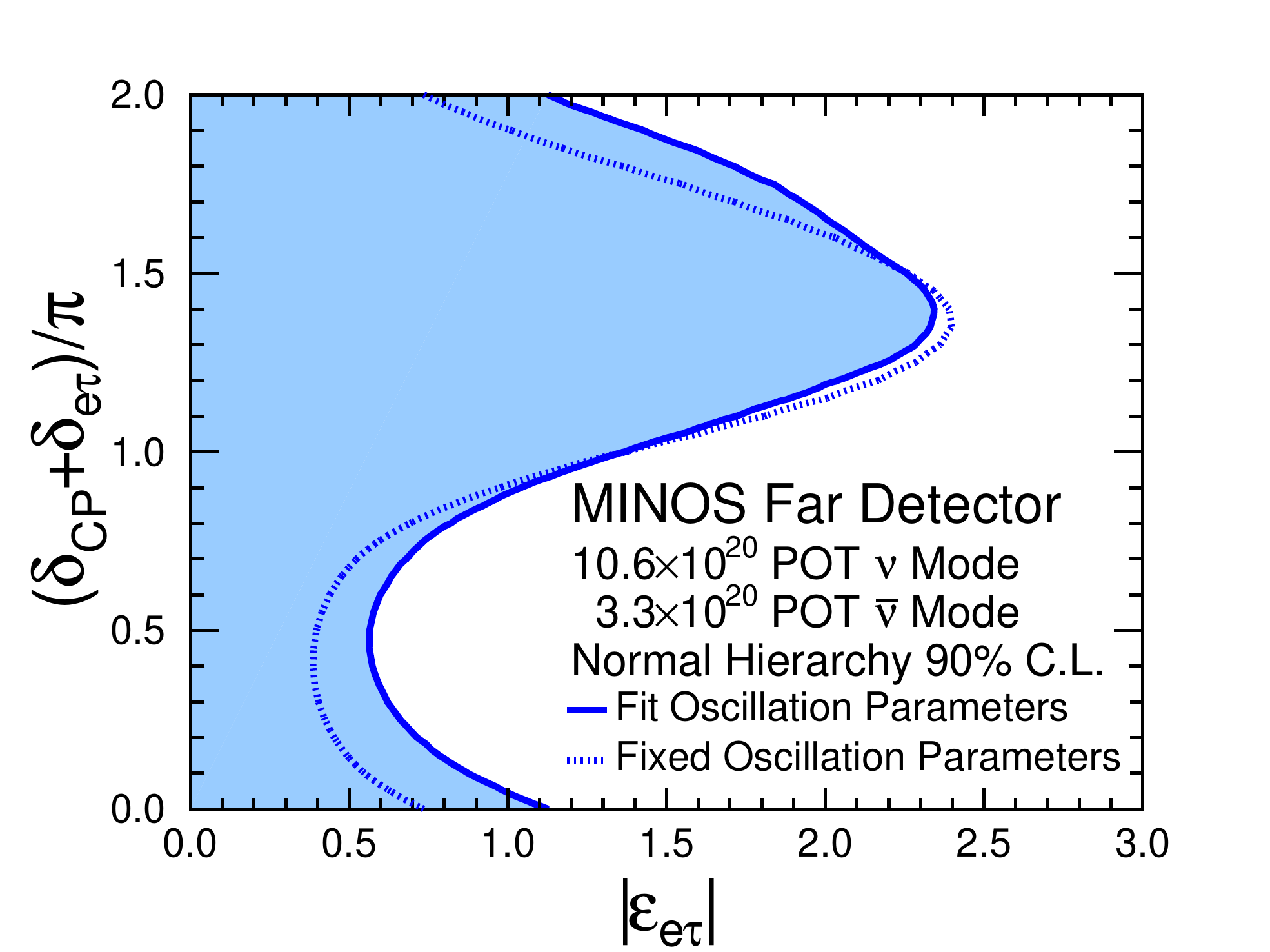}\\
   \includegraphics[trim = 0mm 0mm 0mm 12mm, clip, width=\columnwidth]{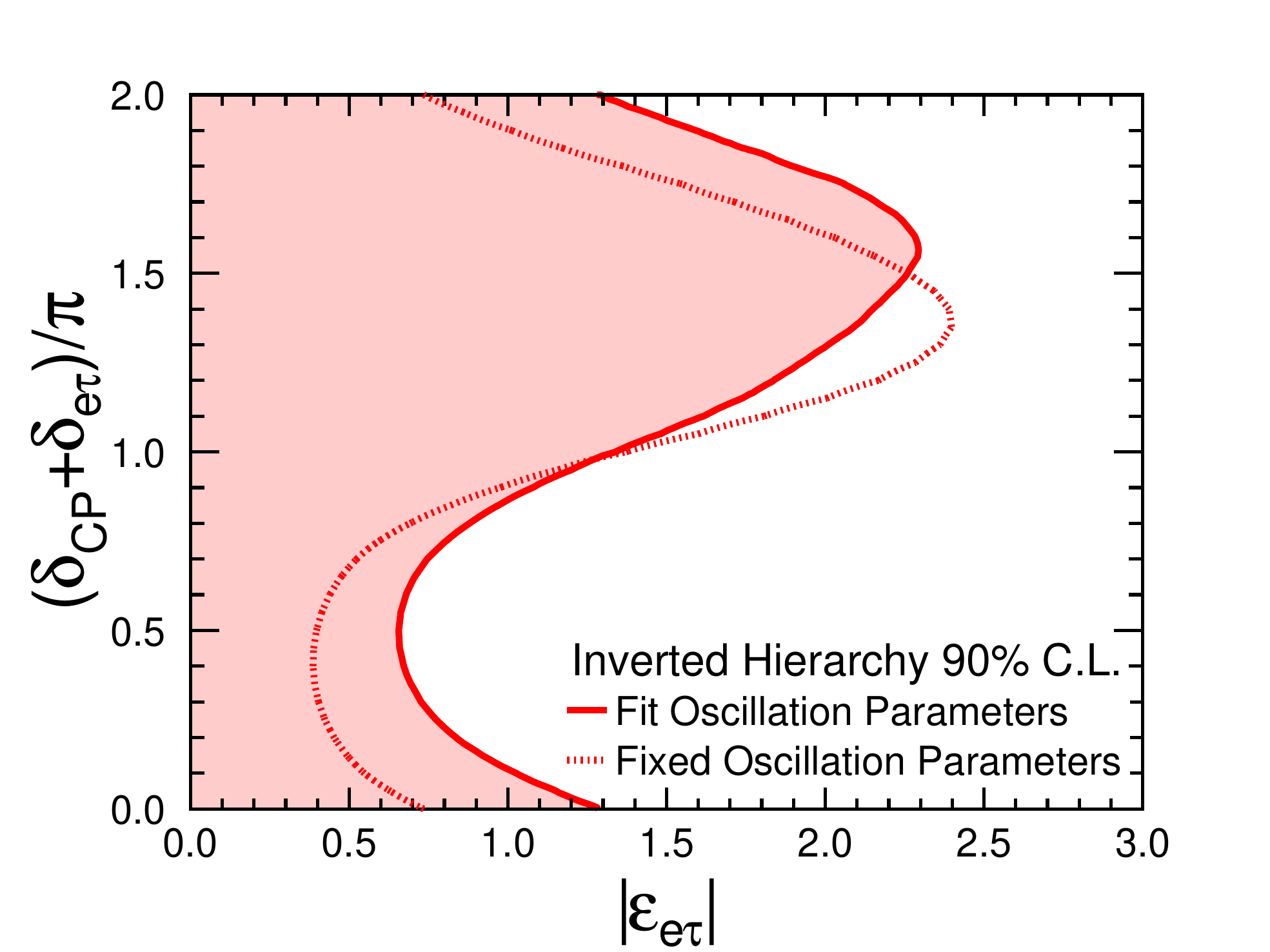}
   \caption{90$\%$ C.L. in the $\magepset,\delcom$ parameter space for normal (top) and inverted (bottom) neutrino mass hierarchy using $\nue$ and $\nuebar$ appearance candidates in the MINOS Far Detector. The shaded areas to the left of the solid contours indicate the MINOS allowed regions where additional oscillation parameters, including NSI, were included in the fit. The dotted contours show the limits where the additional oscillation parameters were fixed to the mean values shown in Table~\ref{table:unc}. \label{fig:result}}
   \end{figure} 
 
  The impact of including the nuisance parameters in the fit was investigated. Comparing the solid and dotted contours in Fig.~\ref{fig:result}, we observe that the addition of the nuisance parameters into the fit does not significantly affect the sensitivity of the analysis to the NSI parameters of interest. The standard deviations from the accepted values of these parameters, as well as the best-fit values of $\magepset$ and ($\dcp+\delta_{e\tau})$, are provided in Table~\ref{table:con}.
 
 The shaded regions in Fig.~\ref{fig:result} show the allowed ranges of $\magepset$ given an effective CP phase and choice of neutrino mass hierarchy. In both cases, the allowed region is consistent with predictions of the standard oscillation model. These results yield modest improvement over the model-independent limit established in \cite{ref:big}, which set $|\epset| < 3.0$ for propagation through Earth-like material, and the limits are consistent with the MINOS+T2K result presented in \cite{ref:Mann}. 
 
In summary, we have performed a direct search for non-standard interactions using the full sample of $\nue$ and $\nuebar$ appearance candidates in the MINOS FD. Using a statistical selection algorithm to identify $\nue$ and $\nuebar$ events, we performed a simultaneous, two-dimensional fit to neutrino and antineutrino samples to place limits upon the $\magepset,\delcom$ parameter space. We found no evidence for non-standard neutrino interactions. The results provide improvement on existing constraints from independent models and are comparable to the previous limit established using MINOS and T2K data. 

This work was supported by the U.S. DOE; the United Kingdom STFC; the U.S. NSF; the State and University of Minnesota; and Brazil's FAPESP, CNPq and CAPES.  We are grateful to the Minnesota Department of Natural Resources, the crew of the Soudan Underground Lab, and the personnel of Fermilab for their contribution to this effort. We thank the Texas Advanced Computing Center at The University of Texas at Austin for the provision of computing resources.
%\newpage

\end{document}